\documentclass{article}
\usepackage{spconf,amsmath,graphicx}
\usepackage{amsfonts,amssymb} 
\usepackage{threeparttable}
\usepackage{multirow}
\usepackage{booktabs}
\usepackage{stfloats}

\title{MFCCA:Multi-Frame Cross-Channel attention for multi-speaker ASR in Multi-party meeting scenario}
\name{\begin{tabular}{c}Fan Yu$^1$, Shiliang Zhang, Pengcheng Guo$^1$, Yuhao Liang$^1$, Zhihao Du, Yuxiao Lin$^2$, Lei Xie$^{1*}$\thanks{* Lei Xie is the corresponding author.}\end{tabular}}
\address{
  $^1$Audio, Speech and Language Processing Group (ASLP@NPU), School of Computer Science,\\ Northwestern Polytechnical University, Xi’an, China \\
  $^2$College of Computer Science and Technology, Zhejiang University, Hangzhou, China}

\begin{document}
%
\maketitle

\begin{abstract}

Recently cross-channel attention, which better leverages multi-channel signals from microphone array, has shown promising results in the multi-party meeting scenario.
Cross-channel attention focuses on either learning global correlations between sequences of different channels or exploiting fine-grained channel-wise information effectively at each time step.
Considering the delay of microphone array receiving  sound, we propose a multi-frame cross-channel attention, which models cross-channel information between adjacent frames to exploit the complementarity of both frame-wise and channel-wise knowledge.
Besides, we also propose a multi-layer convolutional mechanism to fuse the multi-channel output and a channel masking strategy to combat the channel number mismatch problem between training and inference.
Experiments on the AliMeeting, a real-world corpus, reveal that our proposed model outperforms single-channel model by 31.7\% and 37.0\% CER reduction on Eval and Test sets. 
Moreover, with comparable model parameters and training data, our proposed model achieves a new SOTA performance on the AliMeeting corpus, as compared with the top ranking systems in the ICASSP2022 M2MeT challenge, a recently held multi-channel multi-speaker ASR challenge.

\end{abstract}

\section{Introduction}
\label{sec:intro}

Multi-speaker automatic speech recognition (ASR) aims to transcribe speech that contains multiple speakers, and hopefully overlapped speech can be correctly transcribed. It is an essential task of rich transcription in multi-party meetings~\cite{fiscus2005rich,fiscus2006rich,fiscus2007rich}. In recent years, with the advances of deep learning, many end-to-end neural multi-speaker ASR approaches have been proposed~\cite{yu2017recognizing,chen2017progressive,kanda2020serialized} and promising results have been achieved on synthetic multi-speaker datasets, e.g., LibriCSS~\cite{ChenContinuous}. However, transcribing real-world meetings is far more challenging with entangled difficulties such as overlapping speech, conversational speaking style, unknown number of speakers, far-field speech signals with noise and reverberation. Recently, two challenges -- Multi-channel Multi-party Meeting Transcription (M2MeT)~\cite{Yu2022M2MeT,Yu2022Summary} and Multimodal Information based Speech Processing (MISP)~\cite{Chen2022misp} -- have made available valuable real-world multi-talker speech datasets to benchmark multi-speaker ASR towards real conditions and applications.

In the real-world applications, microphone array is usually adopted for far-field speech recording scenarios, including those in M2MET and MISP, where beamforming is a common algorithm to leverage spatial information for multi-channel speech enhancement.
With the help of deep neural networks, time-frequency mask-based beamforming~\cite{wang2018supervised,kinoshita2017neural,heymann2016neural,yu2017permutation,heymann2016neural,erdogan2016improved} has shown superior performance in various multi-speaker benchmarks, such as AMI~\cite{mccowan2005ami}, CHiME~\cite{BarkerWVT18,watanabe20b_chime} and M2MeT~\cite{Yu2022M2MeT,Yu2022Summary}.
The mask estimation network needs to be trained with signal-level criteria on the simulated data where the reference speech is required. Simulated data has a clear gap with real-world data, and optimizing the signal-level criteria may not necessarily lead to lowered word error rate (WER) as well.
Aiming to alleviate such mismatch, joint optimization of multi-channel front-end and ASR has been proposed~\cite{kanda2018hitachi,kanda2019acoustic,subramanian2019speech,subramanian2020far,zhang2020end,chang2020end}.
Under the joint learning framework, the whole system can be optimized with an ultimate ASR loss function by adopting real-world data without reference-cleaned signals.

The \textit{attention} mechanism has been recently introduced to neural beamforming~\cite{chang2020end,tolooshams2020channel}, which performs recursive \textit{non-linear} beamforming on the data represented in a latent space.
Specifically, \textit{cross-channel attention} has been proposed to directly leverage multi-channel signals in a neural speech recognition system~\cite{chang2021end,chang2021multi}. Impressively, such an approach can bypass the complicated front-end formalization and integrate beamforming and acoustic modeling into an end-to-end neural solution. 
This cross-channel attention approach takes the frame-wise multi-channel signal as input and learns global correlations between sequences of different channels,
which can be easily depicted as mapping each channel representation (query) with a set of channel-average representation (key-value) pairs to an output~\cite{chang2021end,chang2021multi}, namely \textbf{f}rame-\textbf{l}evel \textbf{c}ross-\textbf{c}hannel \textbf{a}ttention (\textbf{FLCCA}).
Meanwhile, \textbf{c}hannel-\textbf{l}evel \textbf{c}ross-\textbf{c}hannel (\textbf{CLCCA}) attention has recently achieved remarkable performance on speech separation~\cite{wang2020neural,wang2021continuous} and speaker diarization~\cite{horiguchi2022multi,wang2022cross} tasks, even leading a system to win the first place in the speaker diarization track in M2MeT challenge~\cite{wang2022cross}. 
Compared with FLCCA,  CLCCA is computed along the channel dimension, the representations of each channel are combined with those of the other channels for each time step~\cite{wang2020neural}, which functions similarly as beamforming. 


From our point of view, FLCCA and CLCCA can be complementary in capturing temporal and spatial information. Frame-level is less capable of extracting fine-grained channel-wise patterns since averaging the channel representations directly may deteriorate the individual channel information. Channel-level cross-channel attention, on the other hand, only focuses on leveraging spatial diversities and capturing inter-channel correlations on each time step, without considering the context relationship between different channels.
Thus, in this paper, we exploit the complementarity between frame-level and channel-level cross-channel attention and propose a \textbf{m}ulti-\textbf{f}rame \textbf{c}ross-\textbf{c}hannel \textbf{a}ttention (\textbf{MFCCA}) by modeling both channel-wise and frame-wise information simultaneously.
Direction of arrival (DOA) estimation~\cite{heymann2018frame} has been widely used for speech enhancement, which utilizes the delay of microphone array receiving the signal to estimate the sound source direction based on the phase difference.
Inspired by the intuitive idea behind DOA, our proposed method will pay more attention to channel context between adjacent frames to model both frame-wise and channel-wise dependencies.

We build our MFCCA based multi-channel ASR within an attention based encoder-decoder (AED) structure~\cite{DBLP:journals/corr/VaswaniSPUJGKP17}. 
Moreover, the multi-channel outputs from the encoder are aggregated by multi-layer convolution to reduce channel dimensions gradually. 
Although the cross-channel attention is independent of the number and geometry of microphones, it has the well-known performance degradation issue when number of microphones is reduced~\cite{horiguchi2022multi,wang2020neural}. In order to combat this issue, we propose a channel masking strategy. By randomly masking several channels from the original multi-channel input during training, our MFCCA approach becomes more stable and robust to the arbitrary number of channels.


To the best of our knowledge, we are the first to leverage cross-channel attention on a real meeting corpus -- AliMeeting -- to examine its ability in multi-speaker ASR in meeting scenarios. Experiments on the AliMeeting corpus show that our proposed multi-channel multi-speaker ASR model outperforms the single-channel multi-speaker ASR model by 31.7\% and 37.0\% relative CER reduction on Eval and Test sets, respectively. Moreover, with comparable model parameters and amount of the training data, our proposed model achieves 16.1\% and 17.5\% CER on Eval and Test sets, which surpasses the best system in the M2MeT challenge, resulting in a new SOTA performance on the AliMeeting corpus.


\begin{figure*}[t]
	\centering
	\includegraphics[width=0.95\linewidth]{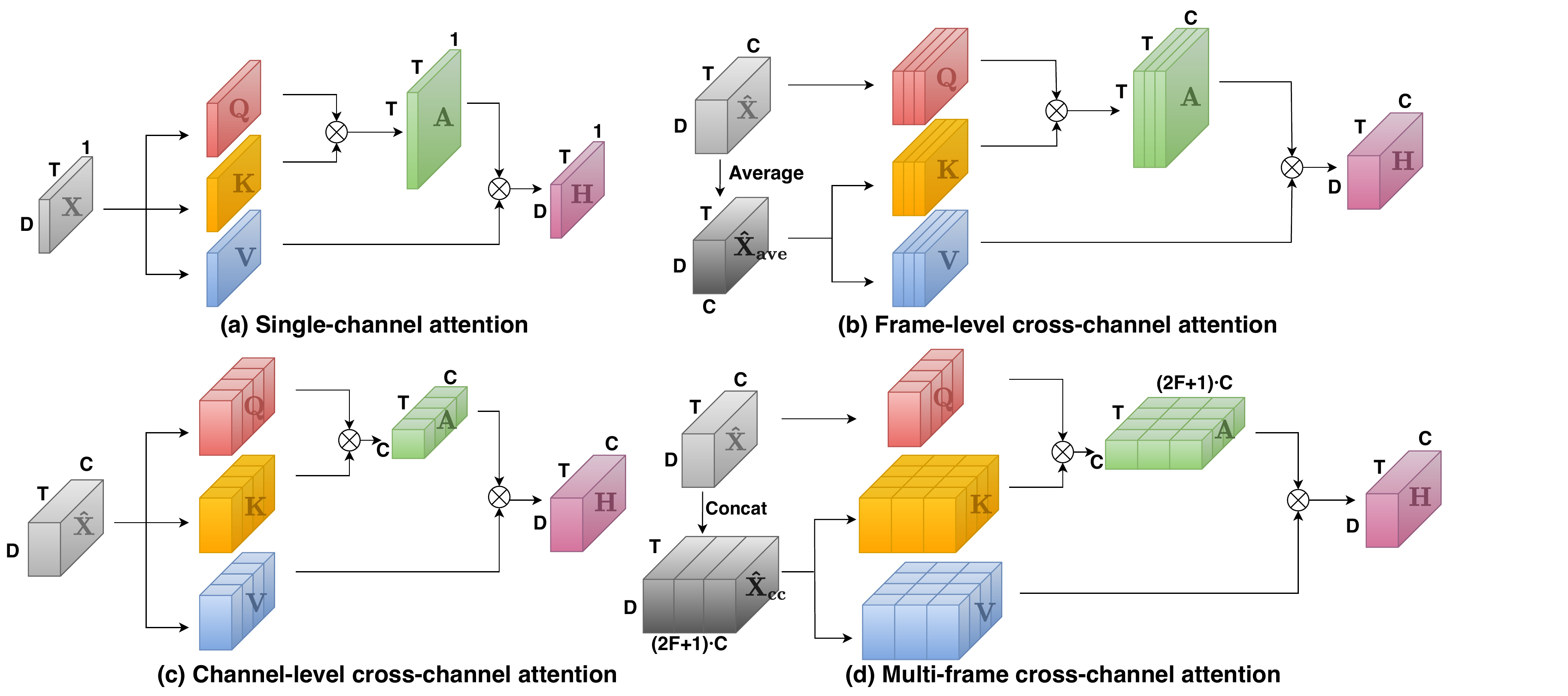}
	\vspace{-0.3cm}
	\caption{
	 Illustration of different attention blocks: (a) Single-channel attention. (b) Frame-level cross-channel attention (FLCCA). (c) Channel-level cross-channel attention (CLCCA). (d) Multi-frame cross-channel attention (MFCCA, proposed).
	}
	\label{attention_layer}
	\vspace{-0.5cm}
\end{figure*}

\section{from single-channel to cross-channel Attention}
\label{sec:attention}
In this section, we first review the multi-headed self-attention commonly used in signal channel cases and then introduce the frame-level and channel-level cross-channel attentions, respectively.
A single channel feature input is defined as $\mathbf{X}$, while a $C$-channel input is formulated as $\bar{\mathbf{X}} = [\mathbf{X}_0, \cdots, \mathbf{X}_{C - 1}]$.
\subsection{Single-channel attention }
\vspace{-0.1cm}

Single-channel attention, which is a standard self-attention structure, adopts the multi-headed scaled dot-product to learn the contextual information within a single channel of speech signal, as shown in Fig.~\ref{attention_layer}a. The output of a single-channel attention for the $i$-th head is calculated as

 \begin{equation}
 \begin{aligned}
    \mathbf{Q}^{sc}_i & =\mathbf{X}\mathbf{W}^{sc,q}_i+(\mathbf{b}^{sc,q}_i)^T \in \mathbb{R}^{ T \times D},  \\
    \mathbf{K}^{sc}_i & =\mathbf{X}\mathbf{W}^{sc,k}_i+(\mathbf{b}^{sc,k}_i)^T \in \mathbb{R}^{ T \times D }, \\
    \mathbf{V}^{sc}_i & =\mathbf{X}\mathbf{W}^{sc,v}_i+(\mathbf{b}^{sc,v}_i)^T \in \mathbb{R}^{ T \times D }, \\
    \mathbf{H}^{sc}_i & = \text{Softmax} \left(\frac{\mathbf{Q}^{sc}_i(\mathbf{K}^{sc}_i)^T}{\sqrt{D}} \right) \mathbf{V}^{sc}_i \in \mathbb{R}^{ T \times D}, \\
 \end{aligned}
 \end{equation}
where $\text{Softmax}(\cdot)$ is the column-wise softmax function, $\mathbf{W}^{sc,*}_i$ and $\mathbf{b}^{sc,*}_i$ are learnable weight and bias parameters for the $i$-th head respectively.

\subsection{Frame-level cross-channel attention }

Frame-level cross-channel attention~\cite{chang2021end,chang2021multi} learns not only the contextual information between time frames but also spatial information across channels, as shown in Fig.~\ref{attention_layer}b. The $i$-th head of FLCCA is calculated as
 \begin{equation}
 \begin{aligned}
 \mathbf{Q}^{fl}_i & =\mathbf{\bar{X}}\mathbf{W}^{fl,q}_i +(\mathbf{b}_i^{fl,q})^T \in \mathbb{R}^{C \times T \times D},\\
 \mathbf{K}^{fl}_i & =\mathbf{\bar{X}}'\mathbf{W}^{fl,k}_i+(\mathbf{b}_i^{fl,k})^T \in \mathbb{R}^{ C \times T \times D },\\
 \mathbf{V}^{fl}_i & =\mathbf{\bar{X}}'\mathbf{W}^{fl,v}_i+(\mathbf{b}_i^{fl,v})^T \in \mathbb{R}^{C \times T \times D},\\
 \mathbf{H}^{fl}_i &= \text{softmax} \left(\frac{\mathbf{Q}_i^{fl}(\mathbf{K}_i^{fl})^T}{\sqrt{D}} \right) \mathbf{V}_i^{fl} \in \mathbb{R}^{ C \times T \times D}, \\
  \end{aligned}
 \end{equation}
%
$\mathbf{\bar{X}}' = [\mathbf{\bar{X}}'_0, \cdots, \mathbf{\bar{X}}'_{C - 1}]$. $\mathbf{\bar{X}}'_c$ is the average of all channels except for the $c^{th}$ channel, which is calculated by $\mathbf{\bar{X}}'_c = (\sum_{n,n\neq c} \mathbf{\bar{X}}_n) / (C - 1) \in \mathbb{R}^{  T \times D}$. $\mathbf{W}^{fl,*}$ and $\mathbf{b}^{fl,*}$ are learnable weight and bias parameters, respectively.

\subsection{Channel-level cross-channel attention }
Channel-level cross-channel attention focuses on leveraging spatial diversities and capturing inter-channel correlations on each time step, as shown in Fig.~\ref{attention_layer}c. The $i$-th head of CLCCA can be formulated as
 \begin{equation}
 \begin{aligned}
 \mathbf{Q}^{cl}_i & =\mathbf{\bar{X}}\mathbf{W}^{cl,q}_i+(\mathbf{b}_i^{cl,q})^T \in \mathbb{R}^{T \times C \times D}, \\
 \mathbf{K}^{cl}_i & =\mathbf{\bar{X}}\mathbf{W}^{cl,k}_i+(\mathbf{b}_i^{cl,k})^T \in \mathbb{R}^{ T \times C \times D }, \\
 \mathbf{V}^{cl}_i & =\mathbf{\bar{X}}\mathbf{W}^{cl,v}_i+(\mathbf{b}_i^{cl,v})^T \in \mathbb{R}^{T \times C \times D}, \\
 \mathbf{H}^{cl}_i & = \text{softmax} \left(\frac{\mathbf{Q}_i^{cl}(\mathbf{K}_i^{cl})^T}{\sqrt{D}} \right) \mathbf{V}^{cl}_i \in \mathbb{R}^{ T \times C \times D}, \\
 \end{aligned}
   \end{equation}
%
Again, $\mathbf{W}^{cl,*}$ and $\mathbf{b}^{cl,*}$ are learnable weight and bias parameters, respectively.

\vspace{-0.3cm}
\section{PROPOSED METHOD}
\label{sec:methods}
\vspace{-0.2cm}

\subsection{Multi-frame cross-channel attention }\label{sec:3-1}

Based on the discussion of FLCCA and CLCCA, multi-frame cross-channel attention is proposed to exploit the complementarity between frame-level and channel-level information, as shown in Fig.~\ref{attention_layer}d. The $i$-th head of MFCCA is calculated as
 \begin{equation}
 \begin{aligned}
 \mathbf{Q}^{mf}_i & =\mathbf{\bar{X}}\mathbf{W}_i^{mf,q}+(\mathbf{b}_i^{mf,q})^T \in \mathbb{R}^{T \times C \times D},\\
 \mathbf{K}^{mf}_i & =\mathbf{\bar{X}}_{cc}\mathbf{W}_i^{mf,k}+(\mathbf{b}_i^{mf,k})^T \in \mathbb{R}^{T \times (2F+1) \cdot C \times D},\\
 \mathbf{V}^{mf}_i & =\mathbf{\bar{X}}_{cc}\mathbf{W}_i^{mf,v}+(\mathbf{b}_i^{mf,v})^T \in \mathbb{R}^{T \times (2F+1) \cdot C \times D},\\
 \mathbf{H}^{mf}_i &= \text{softmax} \left(\frac{  \mathbf{Q}_i^{mf}(\mathbf{K}_i^{mf})^T}{ \sqrt{D}} \right) \mathbf{V}^{mf}_i \in \mathbb{R}^{ T \times C \times D }, \\
 \end{aligned}
 \end{equation}
%
where $\mathbf{W}^{mf,*}$ and $\mathbf{b}^{mf,*}$ are learnable weight and bias parameters, $\mathbf{\bar{X}_{cc}} = [\mathbf{\bar{X}_{cc}^{0}}, \cdots,\mathbf{\bar{X}_{cc}^{t}}, \cdots, \mathbf{\bar{X}_{cc}^T}]$. 
$\mathbf{\bar{X}}_{cc}^t$ is the concatenation of the context frames, which is calculated by$\mathbf{\bar{X}}_{cc}^t = [\mathbf{\bar{X}}^{t-F},...,\mathbf{\bar{X}}^t,...,\mathbf{\bar{X}}^{t+F}] \in \mathbb{R}^{ (2F+1) \cdot C\times D}$. 
$F$ is the number of the past and future frames to be concatenated at each time step, which is a trade-off between performance and computation cost.
Inspired by the DOA calculation which utilizes the delay of the microphone array to estimate the source direction for speech enhancement, our proposed MFCCA focuses on channel context of adjacent frames to improve the ability of modeling the frame-level and channel-level contextual information together.

\vspace{-0.3cm}
\subsection{Conformer block}
\vspace{-0.2cm}

\begin{figure}[!htb]
	\centering
	\includegraphics[scale=0.6]{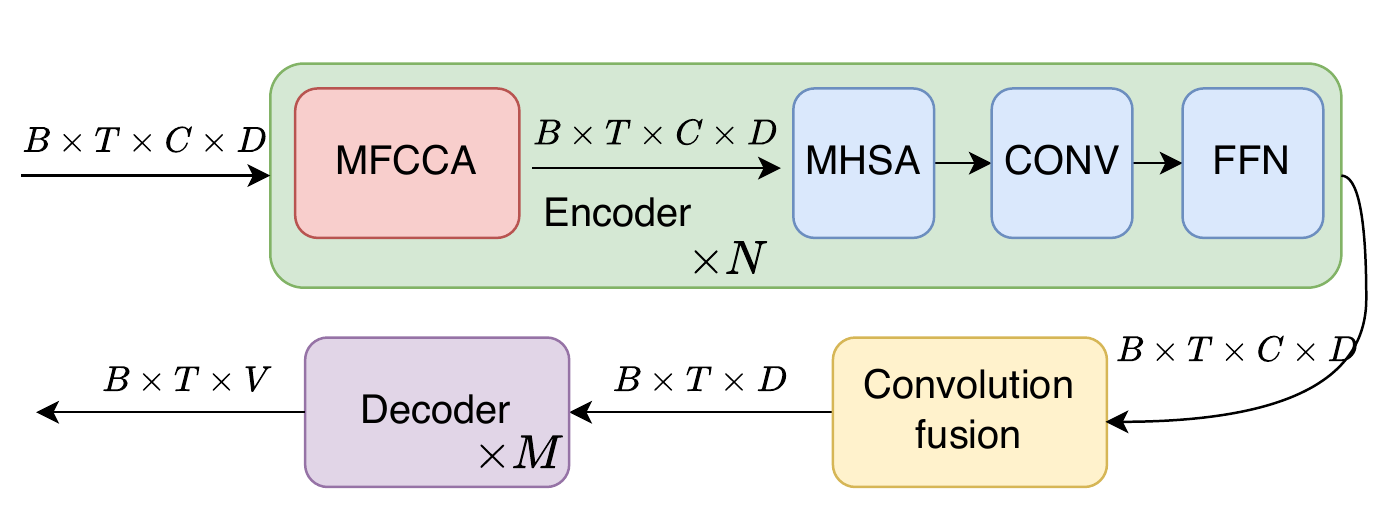}
	\vspace{-5pt}
	\caption{
		An overview of the proposed multi-channel transformer network. 
	}
	\label{model}
\vspace{-18pt}
\end{figure}

Our encoder layer also adopts the Conformer block~\cite{gulati2020conformer, guo2021recent}, which includes a multi-headed self-attention (MHSA) module, a convolution (CONV) module, and a pair of feed-forward  (FFN) module in the Macaron-Net style. 
Conformer models both local and global dependencies of the audio sequence in a parameter-efficient way, which makes full use of the long-range global modeling ability of the MHSA module and the fine-grained local feature extraction ability of the CONV module.
Note that the CONV and FFN module directly follow the multi-frame cross-channel attention will determinate the model performance, which will bring about 1\% absolute CER reduction according to our experiment.
Since the CONV module and FFN module both models at the frame-level, learning of channel dependence by multi-frame cross-channel attention will be affected. Thus, we adopt the model structure in Fig.~\ref{model}.

\vspace{-5pt}

\vspace{-0.2cm}
\subsection{Convolution fusion}
\vspace{-0.1cm}

\begin{figure}[!htb]
	\centering
	\includegraphics[scale=0.45]{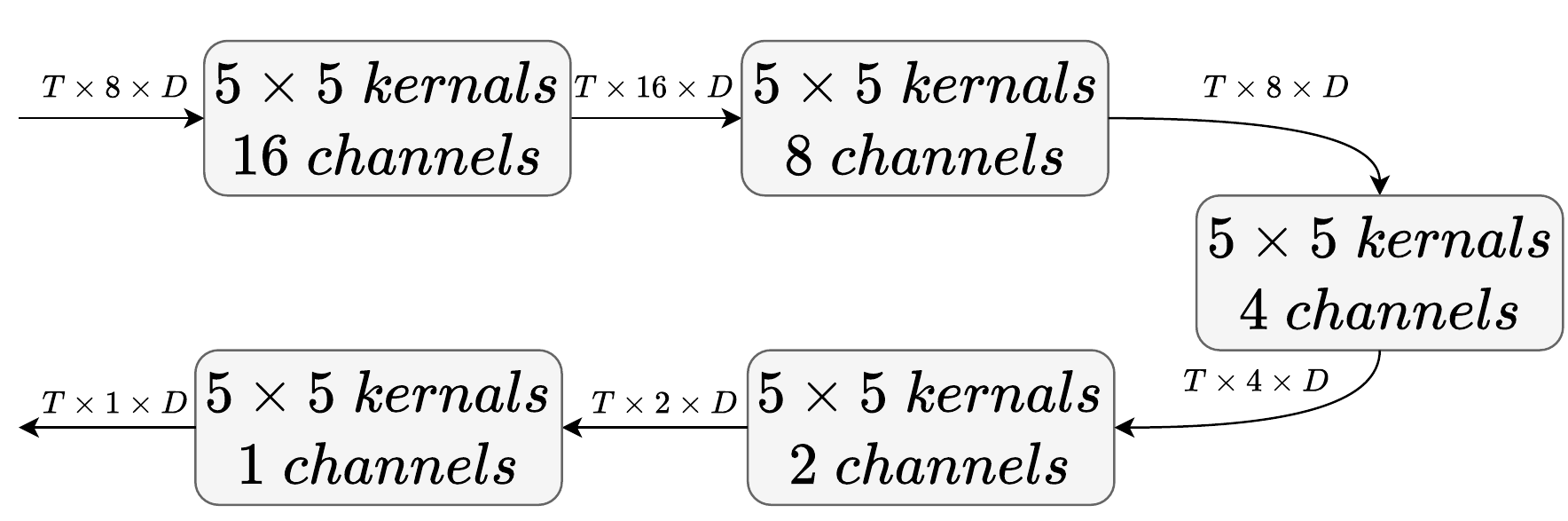}
	\vspace{-5pt}
	\caption{
		The architecture of multi-layer convolution module.
	}
	\label{conv}
\vspace{-15pt}
\end{figure}

To integrate the multi-channel outputs, previous studies~\cite{chang2021multi,wang2022cross} mostly averaged or concatenated channel features along the time axis. In order to mitigate the corruption of channel-specific information caused by reducing the channel dimensions directly, we use a multi-layer convolution module to reduce the channel dimensions gradually. As show in Fig.~\ref{conv}, the multi-layer convolution module consists of five 2-D convolution layers, which only increases negligible parameters. The number of input channels in the multi-layer convolution module is fixed. Therefore, if the channel number of the input is less than the pre-configured value, we need to expand the channel by simple repeating.

\vspace{-5pt}

\vspace{-0.3cm}
\subsection{Channel masking}
\vspace{-0.1cm}

Cross-channel attention is independent of the number of microphones and microphone geometry in its nature. But in practice, the performance of channel-level cross-channel attention is easily affected by the number of the channels~\cite{horiguchi2022multi,wang2020neural}, especially when the channel numbers involved in the decoding and training period are different. Channel dropout~\cite{horiguchi2022multi} was proposed to prevent the models from being overly dependent on spatial information, in which multi-channel inputs are randomly dropped to be a single channel. However, channel dropout mainly improves the speech recognition performance of multi-channel model on a single-channel test set, which does not completely solve the problem of channel number mismatch.
In order to improve the robustness of the model for different channel numbers, we introduce a channel masking strategy, which masks channels randomly for the multi-channel input. Specifically, a uniform probability $p \in (0, 1)$ is used to decide whether the multi-channel input will be masked. 
When choosing to mask, we randomly select $m \in (1,C)$ channels to be masked where $C$ is the total number of channels and $m$ is determined with equal probability $\frac{1}{C}$.
Based on the channel masking strategy, our multi-channel ASR model can easily generalize to variant channel numbers as well as different microphone array geometries, leading to a more practical solution. 

\vspace{-0.3cm}
\subsection{Training strategy}
\vspace{-0.1cm}

Considering the problem of overlapping speech and unknown number of speakers in real-world meeting scenarios, we adopt the Serialized Output Training (SOT) \cite{kanda2020serialized} to enable the multi-speaker recognition ability.
The SOT scheme gets rid of the limitation on the number of speakers and models the dependencies among outputs of different speakers in an effective and simple way.
In the training period, transcriptions of different speakers are serialized into a single word sequence with a special token $\langle \text{sc} \rangle$ inserted. The order of the transcriptions is determined by their start time.
The experiments have shown that the SOT scheme achieves a better CER than the permutation invariant training (PIT) scheme, which needs to calculate all the permutations~\cite{kanda2020serialized}. 

\vspace{-0.4cm}
\section{EXPERIMENTS}
\label{sec:experiment}
\vspace{-0.3cm}

\subsection{Dataset}
\label{ssec:data}
\vspace{-0.1cm}

We use AliMeeting\footnote{http://www.openslr.org/119/} corpus~\cite{Yu2022M2MeT,Yu2022Summary}, a challenging Mandarin meeting dataset with multi-talker conversations, to evaluate our multi-channel multi-speaker ASR model.
The AliMeeting corpus contains 104.75 hours data for training (Train), 4 hours for evaluation (Eval), and 10 hours for test (Test). Each set contains several meeting sessions and each session consists of a 15 to 30 minutes discussion by 2 to 4 participants.
The AliMeeting corpus contains the 8-channel audios recorded from an annular microphone array (\textit{Ali-far}), as well as the near-field audio (\textit{Ali-near}) from the participant's headset microphone. \textit{Ali-far-bf} is produced by applying CDDMA Beamformer~\cite{huang2020differential,zheng2021real}. 
Meanwhile, similar to the M2MeT challenge submissions~\cite{shen2022volcspeech}, we also use the training set of the  \textit{Aishell4}\footnote{http://www.openslr.org/62/}~\cite{fu2021aishell} and 600 hours simulated training data named \textit{Ali-simu} from \textit{Ali-near}, which covers 2-4 speakers in one utterance with 15-40\% overlapping ratio.



\begin{table*}[ht]
\centering
\caption{Results for various multi-channel approaches on Eval and Test sets (\%).}
\begin{threeparttable}[t]

\begin{tabular}{lccccccccccc}
\toprule
\hline
\multicolumn{1}{c}{\multirow{2}{*}{Model}}  & \multicolumn{5}{c}{Eval}         &  & \multicolumn{5}{c}{Test}         \\ \cline{2-6} \cline{8-12} 
\multicolumn{1}{c}{}                        & 1-ch & 2-ch & 4-ch & 6-ch & 8-ch &  & 1-ch & 2-ch & 4-ch & 6-ch & 8-ch \\ \hline
Single channel~\cite{Yu2022M2MeT,Yu2022Summary}                              & 32.3 & 32.3 & 32.3 & 32.3 & 32.3 &  & 33.8 & 33.8 & 33.8 & 33.8 & 33.8 \\
Beamformer~\cite{Yu2022M2MeT,Yu2022Summary}                                  & -    & -    & -    & -    & 30.7 &  & -    & -    & -    & -    & 31.8 \\
Random select                               & 30.2 & 30.2 & 30.2 & 30.2 & 30.2 &  & 31.2 & 31.2 & 31.2 & 31.2 & 31.2 \\
Complex convolution                         & 56.3 & 35.8 & 33.0 & 32.4 & 30.1 &  & 55.6 & 38.4 & 34.7 & 32.4 & 31.0 \\ \hline
Frame-level cross-channel~\cite{chang2021end,chang2021multi}\tnote{$\dagger$}           & 60.5 & 50.4 & 25.9 & 22.6 & 22.5 &  & 63.8 & 51.8 & 27.5 & 24.6 & 24.6 \\
Channel-level cross-channel~\cite{horiguchi2022multi,wang2022cross}\tnote{$\dagger$}         & 38.4 & 27.7 & 21.5 & 20.8 & 20.6 &  & 39.3 & 29.3 & 23.2 & 22.7 & 22.4 \\
Frame-level co-attention~\cite{horiguchi2022multi}\tnote{$\dagger$}                                 & 38.1 & 26.3 & 23.2 & 22.7 & 22.5 &  & 39.1 & 27.9 & 24.4 & 24.2 & 24.0 \\ \hline
Multi-frame cross-channel  & 38.0 & 27.3 & 21.2 & 20.6 & 20.2 &  & 39.0 & 28.8 & 22.9 & 22.3 & 22.0 \\
\quad+ Convolution fusion                        & 37.8 & 26.9 & 20.8 & 20.1 & 19.9 &  & 38.8 & 28.5 & 22.6 & 22.1 & 21.8 \\
\quad  \quad+ Mask channel ($p$=10\%)                      & 36.1 & 25.8 & 20.3 & 19.7 & 19.6 &  & 37.2 & 27.6 & 22.2 & 21.8 & 21.5 \\
\quad  \quad+ Mask channel ($p$=15\%)                      & 35.5 & 25.5 & 20.0 & 19.5 & 19.4 &  & 36.8 & 27.3 & 22.2 & 21.6 & 21.4 \\
\quad  \quad+ Mask channel ($p$=20\%)                      & \textbf{ 35.1} & 25.4 & \textbf{ 20.0} & \textbf{ 19.5} & \textbf{ 19.4} &  & \textbf{ 36.3} & \textbf{ 26.9} &\textbf{  22.0} & \textbf{ 21.5} & \textbf{ 21.3} \\
\quad  \quad+ Mask channel ($p$=25\%)                      & 35.2 & \textbf{ 25.3} & 20.2 & 19.6 & 19.5 &  & 36.6 & 27.7 & 22.1 & 21.6 & 21.4 \\ \hline
\bottomrule
\end{tabular}
\begin{tablenotes}
    \footnotesize
		\item $\dagger$: This models is re-implemented by ourselves with the same parameter structure as our model for fair comparison.
\end{tablenotes}
\end{threeparttable}
\vspace{-0.5cm}

\label{tab:main_result}
\end{table*}

\vspace{-0.3cm}
\subsection{Baselines}
\label{ssec:baseline}
\vspace{-0.2cm}

We compare our MFCCA based multi-channel multi-speaker ASR model with four baselines:
(1) \textit{Single channel model}: as the single channel baseline. Specifically, we use the first channel of \textit{Train-Ali-far} for training and testing.
(2) \textit{Beamformer}: the CDDMA Beamformer~\cite{huang2020differential,zheng2021real} has shown promising results in speech enhancement and it uses all the channels for beamforming, which generates enhanced single channel data (\textit{Ali-far-bf}) for the ASR model.
(3) \textit{Random selection}:  a dynamic strategy is adopted to randomly select a channel of \textit{Train-Ali-far} as the input to the ASR model during training. Note that the first channel is selected as the input for testing.
(4) \textit{Complex convolution}: the multi-channel real and imaginary parts of Short-Time Fourier Transform (STFT) results are extracted for complex convolution~\cite{hu2020dccrn}. The convolution structure is similar to that in Fig.~\ref{conv}.


\vspace{-0.3cm}
\subsection{Experimental setup}
\label{ssec:setup}
\vspace{-0.2cm}

In all experiments, we use the 80-dimensional Mel-filterbank feature extracted with a 25 ms frame length and a 10 ms window shift. The ESPnet~\cite{watanabe2018espnet} toolkit is used to build all our ASR systems. 
We follow the standard configuration of ESPnet to train the baseline models, which contain a 12-layer encoder and 6-layer decoder. The dimension of MHSA and FFN layers are set to 256 and 2048, respectively. For the cross-channel based models, we use an 11-layer encoder and a 6-layer decoder with the 4-head MHSA instead, in order to achieve a similar parameter size to the baseline models.
All the ASR models are trained for 100 epochs and a warmup of the learning rate is used for the first 25,000 iterations. We use 4950 commonly used Mandarin characters as the modeling units. 
Results of all the experiments are measured by Character Error Rate (CER).

\vspace{-0.2cm}
\subsection{Comparison of different multi-channel models}
\vspace{-0.1cm}

As shown in Table~\ref{tab:main_result}, our proposed MFCCA model outperforms the four baselines, especially for the single channel model, leading to 31.7\% (32.3\%$\to$19.4\%) and 37.0\% (33.8\%$\to$21.3\%) relative CER reduction on 8-ch Eval and Test sets, respectively.
Compared with other multi-channel attention models, our MFCCA model shows superior performance, achieving the lowest CER of 20.2\% and 22.0\% on 8-ch Eval and Test sets. When incorporating with the multi-layer convolution fusion to integrate multiple channels, we can obtain further improvement, decreasing the CER from 20.2\%/22.0\% to 19.9\%/21.8\% on 8-ch Eval and Test sets, respectively.

Cross-channel attention models perform well when the channel number of test set is large, but degrade significantly when the number of channels is reduced, e.g., single channel and 2-ch Test sets.
Channel masking can improve the robustness of the model with different channel setups. According to the results, our model obtains the best results on most test sets when channel masking probability is set to 20\%, achieving 7.1\% (37.8\%$\to$35.1\%) and 6.4\% (38.8\%$\to$36.3\%) relative CER reduction on 1-ch Eval and Test sets. 
Meanwhile, channel masking also improves the multi-channel test sets and achieves CERs of 19.4\% and 21.3\% on 8-ch Eval and Test sets, which even has surpassed most of the submissions in the M2MeT challenge~\cite{Yu2022M2MeT,Yu2022Summary}.


\vspace{-0.2cm}
\subsection{Impact of the context frame number}
\vspace{-0.1cm}

As shown in Table~\ref{tab:fram_num}, $F$ is the number of frames that looks back to the past and looks ahead to the future at each time step. 
When increasing $F$ from 0 to 2, we observe that the CER is improved from 20.6\% to 20.0\% on Eval set and 22.4\% to 22.0\% on Test set. 
When further increasing the $F$ from 2 to 4, the gain is marginal on Eval and Test sets, which only brings 0.1\% absolute CER reduction on Eval set.
The reason might be that the channel information of adjacent frames is more important in cross-channel attention, which denotes the importance of the delay time between microphones. Based on this conclusion, the frame number for looking back and ahead is set to 2 in the remaining experiments.

\begin{table}[!htb]
\vspace{-0.4cm}
\caption{Results of MFCCA model with different context frame number on Eval and Test sets (\%).}
\centering
\setlength{\tabcolsep}{11pt}

\begin{tabular}{cccccc}
\toprule

\hline
F    & 0    & 1    & 2    & 3    & 4    \\ \hline
Eval & 20.6 & 20.4 & 20.2 & 20.2 & \textbf{20.1} \\ 
Test & 22.4 & 22.1 & \textbf{22.0} & 22.0 & 22.0 \\ \hline
\bottomrule
\end{tabular}
\label{tab:fram_num}
\vspace{-0.3cm}

\end{table}

\vspace{-0.2cm}
\subsection{Visualization of MFCCA scores}
\begin{figure}[!htb]

	\centering
	
	\includegraphics[scale=0.47]{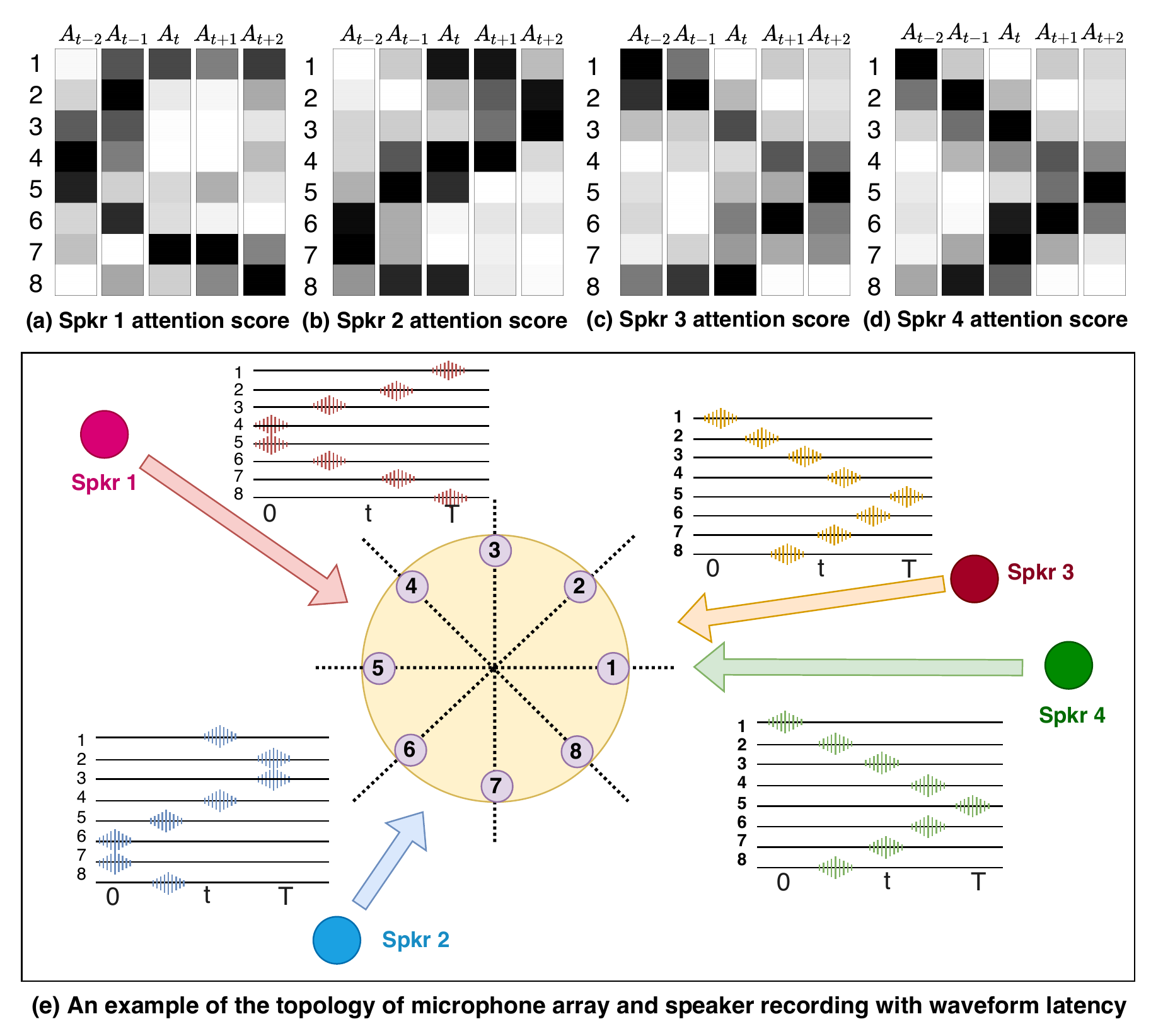}
	\vspace{-10pt}
	\caption{
		Illustration of: (a-d) the attention scores of different speakers. (e) An example of the topology of microphone array and the recorded 8-ch waveforms.
	}
	\label{attention_score}
\vspace{-0.2cm}
\end{figure}

\vspace{-0.1cm}
To analyze the behavior of our proposed model, Fig.~\ref{attention_score} visualizes the attention scores of our MFCCA module and the detailed recording process of the microphone array.
As shown in Fig.~\ref{attention_score}(e), the different microphone-speaker distances may result in time delays during the recording.
For example, the 7-th channel of speaker-1 shows a slight time delay compared with the 4-th channel, since the 4-th microphone is much closer to the speaker.
Fig.~\ref{attention_score} (a-d) are heatmaps of the averaged attention scores computed by our MFCCA module for different speakers.
As described in~\ref{sec:3-1}, for a specific time $t$, its input feature will be appended with two past and future contexts, and the MFCCA module tries to exploit cross-channel dependencies between adjacent frames. 
Combining Fig.~\ref{attention_score}(a) and Fig.~\ref{attention_score}(e), we can find that our model indeed captures the microphone delay information like beamforming, as the model attends more on the 4-th/5-th channels at time $t-2$ and 7-th channel at time $t$.
Note that the attention scores are from the first encoder layer, in which each channel has not yet integrated the other channel information.


\vspace{-0.4cm}
\subsection{Impact of the different training data scale}
\vspace{-0.1cm}


As shown in Table~\ref{tab:big_data}, we compare the results of our proposed model trained with different data scales on Eval and Test sets.
In order to strengthen the acoustic modeling ability of the model, we include the \textit{Train-Ali-near} and \textit{Aishell4} sets into our training, which yields 10.8\% (19.4\%$\to$17.3\%) and 13.6\% (21.3\%$\to$18.4\%) relative CER reductions on Eval and Test sets, respectively.
We also simulate 600 hours 8-channel meeting data based on the \textit{Train-Ali-near} to have a fair comparison with the M2MeT challenge submissions. By using the same simulated data augmentation strategy, our model obtains further improvement, achieving 16.5\% and 18.0\% CERs on Eval and Test sets. 
Meanwhile, we also integrate a neural network language model (NNLM) into our proposed model to improve the language generalization ability, which brings 2.4\% (16.5\%$\to$16.1\%) and 2.7\% (18.0\%$\to$17.5\%) relative CER reductions on Eval and Test sets.
The NNLM is trained on the transcriptions of training data, using extra text data is prohibited according to the M2MeT challenge rule.
Compared with the submission system of the $2^{nd}$ ranking team in M2MeT, which adopted the front-end and back-end joint modeling scheme~\cite{Yu2022Summary,shen2022volcspeech}, our proposed MFCCA model brings 16.1\% (19.2\%$\to$16.1\%) and 15.9\% (20.8\%$\to$17.5\%) relative CER reductions on Eval and Test sets, while the parameters and training data are at a comparable scale.
Furthermore, our model even outperforms the large model of the $1^{st}$ ranking team's submission system~\cite{Yu2022Summary,ye2022royalflush} trained on a large data scale by data augmentation and simulation, leading to 8.0\% (17.5\%$\to$16.1\%) and 6.9\% (18.8\%$\to$17.5\%) relative CER reductions on Eval and Test sets, respectively.

\begin{table}[!htb]
\vspace{-0.3cm}

\caption{Results of MFCCA model with the different training data scales on Eval and Test sets (\%).}
\centering
\setlength{\tabcolsep}{2pt}

\begin{threeparttable}[t]

\begin{tabular}{lcccc}
\toprule

\hline
Model & Para(M) & Data(hrs) & Eval & Test \\ \hline
$1^{st}$ranking w/ model fusion\cite{ye2022royalflush}                     & 114 & 14,000       & 17.5 & 18.8 \\
$1^{st}$ranking~\cite{ye2022royalflush}                     & 114 & 10,000       & 19.1 & 20.1 \\
$2^{nd}$ranking~\cite{shen2022volcspeech}                     & 48 & 917        & 19.2 & 20.8 \\\hline
MFCCA (\textit{Train-Ali-far})          & 45 & 105        & 19.4 & 21.3 \\
\quad  + \textit{Train-Ali-near}, \textit{Aishell4} & 45 & 317        & 17.3 & 18.4 \\
\quad \quad  + \textit{Ali-simu}                  & 45 & 917        &  16.5    & 18.0     \\ 
\quad \quad \quad + NNLM  & 45 & 917        &  \textbf{16.1}    &  \textbf{17.5}    \\ \hline
\bottomrule

\end{tabular}
\end{threeparttable}
\label{tab:big_data}
 \vspace{-0.5cm}

\end{table}

\vspace{-0.2cm}
\section{CONCLUSIONS}
\label{sec:conclude}
\vspace{-0.2cm}

In this work, we propose a multi-frame cross-channel attention (MFCCA) module based on the multi-speaker SOT framework to capture both temporal and spatial information, which exploits the complementarity between frame-level and channel-level cross-channel attention.
Considering the delay of microphone array receiving sound, our MFCCA approach models cross-channel information between adjacent frames.
Besides, we also propose a multi-layer convolutional mechanism to fuse the multi-channel output efficiently.
Finally, in order to combat the channel number mismatch problem between training and inference, we propose a channel masking strategy to improve the robustness of the model with different channel setups.
Evaluated on the real meeting corpus AliMeeting, our proposed model outperforms single channel ASR model by 31.7\% and 37.0\% relative CER reductions on Eval and Test sets, respectively. 
Moreover, with the comparable model parameters and training data, our proposed model achieves a SOTA error rate compared with top ranking systems in the ICASSP2022 M2MeT challenge, the recently held multi-channel multi-speaker ASR challenge.
In the future, we would like to integrate our proposed multi-channel multi-speaker model into speaker-attributed automatic speech recognition for real-world applications. 

\vspace{-0.2cm}
\section{Acknowledgement}
\vspace{-0.2cm}

This work was supported by Alibaba Group through Alibaba Research Intern Program and Key R \& D Projects of the Ministry of Science and Technology (2020YFC0832500).

\bibliographystyle{IEEEbib}
\bibliography{strings,refs}

\end{document}